\documentclass[a4paper,12pt]{article}
\usepackage{amsmath}
\usepackage{amsthm}
\usepackage{amssymb}
\usepackage{mathrsfs}
\usepackage{graphicx}
\usepackage{cite}
\usepackage{indentfirst}
\usepackage{braket}
\usepackage{hyperref}
\hypersetup{hidelinks,colorlinks=false}

\theoremstyle{definition}
\theoremstyle{plain}

\theoremstyle{remark}
\newtheorem*{remark}{Remark}

\DeclareMathOperator{\Tr}{Tr} %\DeclareMathOperator{\Ext}{Ext}
\DeclareMathOperator{\TrB}{Tr_B}

\DeclareMathOperator{\spec}{spec}

\renewcommand*{\Re}{\mathop{\mathrm{Re}}\nolimits}
\renewcommand*{\Im}{\mathop{\mathrm{Im}}\nolimits}
\renewcommand*{\spec}{\mathop{\mathrm{spec}}\nolimits}

\newcommand*{\ZS}{Z_{\rm S}}
\newcommand*{\ZB}{Z_{\rm B}}
\newcommand*{\ZSB}{Z_{\rm SB}}
\newcommand*{\ZMF}{Z_{\rm MF}}
\newcommand*{\HS}{H_{\rm S}}
\newcommand*{\HB}{H_{\rm B}}
\newcommand*{\HI}{H_{\rm I}}
\newcommand*{\HMF}{H_{\rm MF}}
\newcommand*{\rhoSb}{\rho_{{\rm S},\beta}}
\newcommand*{\rhoBb}{\rho_{{\rm B},\beta}}

\begin{document}

\author{Grigorii Timofeev${}^1$
 and 
Anton Trushechkin${}^2$\footnote{e-mails: timofeev.gm@phystech.edu, trushechkin@mi-ras.ru}}

\title{\Large\textbf{Hamiltonian of mean force in the weak-coupling and high-temperature approximations and refined quantum master equations}}

\date{\textit{\small 
${}^1$Moscow Institute of Physics and Technology\\ Institutskii Pereulok 9,
141700 Dolgoprudny, Moscow Region, Russia \\[6pt]
${}^2$Steklov Mathematical Institute of Russian Academy of Sciences\\ Gubkina St. 8, 119991 Moscow, Russia
}}

\maketitle

\begin{abstract}
The Hamiltonian of mean force is a widely used concept to describe the modification of the usual canonical Gibbs state for a quantum system whose coupling strength with the thermal bath is non-negligible. Here we perturbatively derive general approximate expressions for the Hamiltonians of mean force in the weak-coupling approximation and in the high-temperature one. We numerically analyse the accuracy of the corresponding expressions and show that the precision of the Bloch-Redfield equantum master equation can be improved if we replace the original system Hamiltonian by the Hamiltonian of mean force.

%The Hamiltonian of mean force is a widely used concept to describe the equilibrium state of a quantum system in contact with a thermal bath. Non-negligible interactions between the system and the bath lead to corrections to the usual canonical Gibbs state. This modified equilibrium state can be expressed as the Gibbs state with respect to the modified system Hamiltonian -- Hamiltonian of mean force, which takes into account non-negligible coupling to the bath. Hamiltonian of mean force is used in quantum thermodynamics and quantum master equations. 

\end{abstract}

\section{Introduction}

Thermalization under the influence of a thermal bath is a fundamental property of open quantum systems. When the coupling to the bath is treated as negligibly small (sometimes this regime is referred to as the ultraweak coupling regime \cite{TMCA}), then the reduced equilibrium state of the system approaches the canonical Gibbs state with the same temperature as bath's temperature. It is fully specified by the system Hamiltonian and does not depend on the bath and details of the system-bath interaction. However, there is a growing interest to corrections to the canonical Gibbs state originating from non-negligible coupling strength, see Ref.~\cite{TMCA} for a review. 

Exact expression for the equilibrium state of the system for an arbitrary coupling strength is known only for particular cases, e.g., for an oscillator coupled to a thermal bath of harmonic oscillators \cite{Grabert1984} or, more generally, for a system of a finite number of harmonic oscillators coupled to a thermal bath of harmonic oscillators \cite{Subasi}. In other cases, perturbation theories can be used.

The weak coupling regime is a widely used approximation, from which a perturbation theory can be developed. In this regime, the system-bath coupling strength is non-negligible, but can be treated perturbatively. The corresponding corrections to the Gibbs state for particular and general systems were derived in Refs.~\cite{Subasi,Geva,Mori,Thingna,Purkaqubit,
CresserAnders}, see also Ref.~\cite{LatuneReact}. Expressions for the equilibrium state of the system at the opposite regime of ultrastrong coupling were derived in Refs.~\cite{CresserAnders,LatuneUltrastrong}. The polaron-transformation approach is one of the ways to interpolate between these two opposite regimes. Perturbative expressions for the equilibrium state of the system in this approximation were derived in Refs.~\cite{Cao2012jcp,Cao2012pre,Cao2016}. Another kind of approximation is the high-temperature approximation. For this regime, a perturbative expression for the equilibrium state in a specific model used in the theory of excitation energy transfer was derived in Ref.~\cite{ValkunasHMF}. It also interpolates between weak- and ultrastrong-coupling regimes whenever the temperature is high enough.

A perturbation theory can be developed for the reduced density operator itself. In the case of weak-coupling limit the corresponding perturbative series is
\begin{equation}\label{EqRhoSbSeries}
\rhoSb=\rhoSb^{(0)}+\lambda^2\rhoSb^{(2)}+
\lambda^4\rhoSb^{(4)}+\ldots,
\end{equation}
where $\rhoSb^{(0)}\propto e^{-\beta\HS}$ is the canonical Gibbs state with $\HS$ being the system Hamiltonian and the subsequent terms are corrections and $\beta$ being the inverse temperature of the bath. Here $\lambda$ is the system-bath coupling strength and, under certain conditions, the terms with the odd powers of $\lambda$ disappear.

However, often, the equilibrium state of the system is expressed in a quasi-Gibbsian form using the concept of the Hamiltonian of mean force $\HMF(\beta)$, which is temperature-dependent: $\rhoSb\propto e^{-\beta\HMF(\beta)}$. The Hamiltonian of mean force and the corresponding partition function $\ZMF=\Tr e^{-\beta\HMF(\beta)}$ are used in strong-coupling quantum thermodynamics \cite{GelinThoss,Jarzynski,MillerAnders,MillerHMF,StrasbergEsposito,
Rivas,TalknerHanggi}. One can also develop a perturbation theory not for the density operator itself but for the Hamiltonian of mean force  (in other words, for the ``generator'' of $\rhoSb$ in imaginary time):
\begin{equation}
\HMF(\beta)=\HS+\lambda^2\HMF^{(2)}(\beta)
+\lambda^4\HMF^{(4)}(\beta)+\ldots
\end{equation}
In this paper, we derive a general expression for the first non-vanishing correction $\HMF^{(2)}(\beta)$ to the Hamiltonian of mean force.\footnote{Independently, the  same result has been obtained by Marcin {\L}obejko, Marek Winczewski, Gerardo Suarez, Micha{\l} Horodecki, and Robert Alicki.}

It is worthwhile to mention paper \cite{TereEffGibbs}, where the effective Hamiltonian with corrections to the Gibbs state in the weak-coupling regime was derived for the time-averaged observables.

Another result of the present paper is a generalization of the high-temperature approximation for the Hamiltonian of mean force obtained in Ref.~\cite{ValkunasHMF} to the  general model of system-bath interactions (for the bath of harmonic oscillators).

The Hamiltonian of mean force is used in Refs.~\cite{KM,KMcorr,MerkliRev} to refine the Markovian quantum master equations for the weak coupling regime. The Gibbs state $\rhoSb^{(0)}\propto e^{-\beta\HS}$ is known to be a steady state of the Davies (or secular Bloch-Redfield) generator \cite{BP,RH}. If we formally replace the system Hamiltonian $\HS$ with the Hamiltonian of mean force $\HMF(\beta)$ in the construction of the master equation, then the steady state of the corresponding generator (which we will call the refined generator) coincides with the exact steady state $\rhoSb$ in this case. However, rigorous estimations obtained in these papers do not guarantee that such master equation does not lose the precision on intermediate time scales. Here we use concrete approximate expressions for the Hamiltonian of mean force to numerically test the performance of the refined Bloch-Redfield quantum master  equation.

It is our pleasure to dedicate this paper to the 75th anniversary of our teacher, Professor Igor Vasil'evich Volovich. Problems of thermalization and irreversible dynamics as well as classical and quantum kinetic equations is one of the topics of his very broad scientific interests 
\cite{AccLuVol,AccPechVol,PechVol,AVK,VolFunc,VolTrush2009,
VolBog,VolPisk,VolFuncStoch,VolHoloTherm,TrushVolLocGlob,InoVol}, where he has made fundamental contributions. In particular, he is one of the inventors of the theory of quantum stochastic limit \cite{AccLuVol,AccPechVol,PechVol,AVK} and the inventor of functional mechanics \cite{VolFunc,VolTrush2009,
VolBog,VolPisk,VolFuncStoch}. The theory of quantum stochastic limit generalizes the theory of quantum master equations and allows one to derive not only a master equation for the reduced state of the system, but an approximate quantum stochastic equation for the whole system-bath complex. The main idea of the functional mechanics is to treat equations of statistical mechanics as fundamental equations even for individual particles to make the dynamics irreversible even in the microscopic level. We would like to wish to Igor Vasil'evich good health and new bright scientific results.

The following text is organized as follows. In Sec.~\ref{SecMFG}, we give known results about corrections to the Gibbs state (in the sense of perturbation series (\ref{EqRhoSbSeries})). In Sec.~\ref{SecHMF}, we derive the second-order (in $\lambda$) contribution to the Hamiltonian of mean force in the weak-coupling approximation. In Sec.~\ref{SecHMFexp}, we derive an asymptotically equivalent  expression for this contribution in the exponential form (i.e., which gives the same result in the second order). In these sections, we formally do not impose any restrictions on the form of the bath. In Sec.~\ref{SecHarmonic}, we apply the obtained results for the case of the bath of harmonic oscillators. In Sec.~\ref{SecHigh}, we derive the high-temperature approximation for the Hamiltonian of mean force. In Sec.~\ref{SecSim}, we perform numerical comparison of various approximations with the exact  $\rhoSb$ (for the case of the ``bath'' containing a single harmonic oscillator). Finally, Sec.~\ref{SecRef} is devoted to the refined quantum master equations and their numerical performance.

\section{Corrections to the Gibbs state}\label{SecMFG}

Consider a system coupled to an environment (``bath''):
\begin{equation}
H=\HS+\HB+\lambda \HI,
\end{equation}
where
%\begin{equation}
%\HB=\int d\xi\, \omega(\xi) a^\dag(\xi)a(\xi),
%\end{equation}
\begin{equation}\label{EqHI}
\HI=\sum_\mu A_\mu\otimes B_\mu 
\end{equation}
and $\lambda$ is a small parameter. For simplicity, without loss of generality, we assume that
$\TrB(B_\mu\rhoBb)=0$ for all $\mu$, where
\begin{equation}
\rhoBb=\ZB^{-1}e^{-\beta\HB},
\quad \ZB=\Tr e^{-\beta\HB},
\end{equation}
is the equilibrium state of the isolated bath. Usually, the bath is considered to be large. Here, it is not important and the bath may be small (i.e., one oscillator or one qubit). Note that, sometimes, the Hamiltonian of an open quantum system includes and additional reorganization term proportional to $\lambda^2$. Often, it naturally appears from physical models and, for the case where both system and bath are infinite-dimensional, is necessary for the thermodynamic stability of the whole system+bath complex for large enough $\lambda$. It nontrivially acts only on the system space, but originates from the interaction with the bath \cite{CresserAnders,ValkunasHMF}. We can consider it to be included into the system Hamiltonian $H_S$ and will not treat it as a perturbative term.

We are interested in the reduced state of the system which corresponds to the ``global'' equilibrium with the inverse temperature $\beta$, i.e.,
\begin{equation}
\rho_{{\rm S},\beta}=\TrB\left(\ZSB^{-1}e^{-\beta H}\right),
\quad Z_{\rm SB}=\Tr e^{-\beta H}.
\end{equation}
If $\lambda=0$, then $e^{-\beta H}=e^{-\beta\HS}\otimes e^{-\beta\HB}$ and $\rho_{{\rm S},\beta}=\rho_{{\rm S},\beta}^{(0)}$, where

\begin{equation}
\rhoSb^{(0)}=\ZS e^{-\beta\HS}, \quad \ZS=\Tr e^{-\beta\HS}.
\end{equation}
We want to find the expression for $\rhoSb$ in the second order in $\lambda$:
\begin{equation}
\rhoSb\cong\rhoSb^{(0)}+\lambda\rhoSb^{(1)}+
\lambda^2\rhoSb^{(2)}.
\end{equation}

%Firstly, let us ``renormalize'' $\HS$ as
%\begin{equation}
%\HS'=H_S+\lambda^2\Delta H_S,
%\end{equation}
%where $\Delta H_S$ will be fixed later. In the simplest case, it is equal to zero.

The desired expression can be found using the Kubo expansion (analogous to the corresponding expansion for the time evolution operator, albeit for imaginary time):
\begin{equation}\label{EqKubo}
e^{-\beta(H_0+\lambda V)}
=e^{-\beta H_0}
\left(
1-\lambda\int_0^\beta d\beta_1\,V(\beta_1)
+\lambda^2
\int_0^\beta d\beta_1
\int_0^{\beta_1} d\beta_2\,
V(\beta_1)V(\beta_2)+\ldots
\right),
\end{equation}
for an arbitrary reference Hamiltonian $H_0$ and a perturbation $V$,
where we use the notation of the  unperturbed evolution in  imaginary time:
\begin{equation}
V(\beta')=e^{\beta' H_0}Ve^{-\beta' H_0}.
\end{equation}
In our case, $H_0=\HS+\HB$ and $V=\HI$. The partial trace over the bath gives 
\begin{equation}\label{EqMFGgen}
\begin{split}
\TrB e^{-\beta H}
&=
\ZB\,e^{-\beta\HS}
\left[
1
+\lambda^2
\int_0^{\beta}d\beta_1
\int_0^{\beta_1}d\beta_2\,
\langle
\HI(\beta_1)\HI(\beta_2)
\rangle_{{\rm B},\beta}
+\ldots
\right]
\\
&=
\ZB\,e^{-\beta\HS}
\left[
1+
\lambda^2\sum_{\mu,\nu}
\int_0^\beta d\beta_1
\int_0^{\beta_1} d\beta_2\,
C_{\mu\nu}(\beta_2)
A_\mu^\dag(\beta_1)A_\nu(\beta_2)
+\ldots
\right],
\end{split}
\end{equation}
where $\langle O\rangle_{{\rm B},\beta}=\TrB(O\rhoBb)$ and
\begin{equation}
C_{\mu\nu}(\beta')=\TrB[B^\dag_\mu(\beta')B_\nu\rhoBb]
=\TrB[e^{\beta'\HB} B^\dag_\mu e^{-\beta'\HB}B_\nu\rhoBb]
\end{equation}
is the bath correlation function in  imaginary time.

To  express $A_\mu(\beta')$, we can use the decomposition
\begin{equation}\label{EqAomega}
A_\mu=\sum_\omega A_{\mu\omega},\qquad 
[\HS,A_{\mu\omega}]=-\omega A_{\mu\omega},
\end{equation}
where the summation is over the Bohr frequencies $\omega$ and $\omega'$ of $\HS$. In other words, $\omega,\omega'\in\spec [H_S,\cdot]$. Explicitly,
\begin{equation}
A_{\mu\omega}=
\sum_{
\begin{smallmatrix}
\varepsilon,\varepsilon'\in\spec\HS,\\
\varepsilon-\varepsilon'=\omega
\end{smallmatrix}
}
P_{\varepsilon'}A_\mu P_{\varepsilon},
\end{equation}
where $P_\varepsilon$ are projectors onto the eigenspaces of $\HS$ corresponding to the eigenvalues $\varepsilon$. Then,
\begin{equation}
A_\mu(\beta')=\sum_\omega A_{\mu\omega}e^{-\beta'\omega}
\end{equation}
and, up to the second order in $\lambda$,
\begin{equation}\label{EqMFGnonnorm}
\TrB e^{-\beta H}\cong\ZB e^{-\beta\HS}
\left[
1+
\lambda^2\sum_{\mu,\nu}\sum_{\omega,\omega'}
G_{\mu\nu}(\omega,\omega')
A_{\mu\omega'}^\dag A_{\nu\omega}
\right],
\end{equation}
 where $A_{\mu\omega'}^\dag\equiv (A_{\mu\omega'})^\dag$ and
\begin{equation}
G_{\mu\nu}(\omega,\omega')=
\int_0^\beta d\beta_1
\int_0^{\beta_1} d\beta_2\,
e^{\beta_1\omega'-\beta_2\omega}\,
C_{\mu\nu}(\beta_1-\beta_2).
\end{equation}
The functions $G_{\mu\nu}(0,0)$ are the lineshape functions known in spectroscopy, albeit, again, in  imaginary time.  Formula (\ref{EqMFGnonnorm}) was obtained in Refs.~\cite{Subasi,Geva,Mori,Thingna,Purkaqubit,
CresserAnders,LatuneReact}.

In order to obtain a density operator with the unit trace from expression (\ref{EqMFGnonnorm}), we should divide this expression by its trace:
\begin{equation}\label{EqMFG}
\rhoSb\cong\ZMF^{-1} \rhoSb^{(0)}
\left[
1+
\lambda^2\sum_{\mu,\nu}\sum_{\omega,\omega'}
G_{\mu\nu}(\omega,\omega')
A_{\mu\omega'}^\dag A_{\nu\omega}
\right],
\end{equation}
where
\begin{equation}
\ZMF=\frac{\ZSB}{\ZS\ZB}\cong
1+
\lambda^2\sum_{\mu,\nu}\sum_{\omega,\omega}
G_{\mu\nu}(\omega,\omega)
\Tr(A_{\mu\omega}^\dag A_{\nu\omega}\rhoSb^{(0)}).
\end{equation}
As in Ref.~\cite{CresserAnders}, we can treat $\ZMF^{-1}$ perturbatively as well and obtain
\begin{equation}
\ZMF^{-1}\cong
1-
\lambda^2\sum_{\mu,\nu}\sum_{\omega,\omega}
G_{\mu\nu}(\omega,\omega)
\Tr(A_{\mu\omega}^\dag A_{\nu\omega}\rhoSb^{(0)})
\end{equation}
and 
\begin{equation}\label{EqMFGjim}
\begin{split}
\rhoSb\cong\rhoSb^{(0)}
\Big\{1&+
\lambda^2\sum_{\mu,\nu}\sum_{\omega}
G_{\mu\nu}(\omega,\omega)
\big[
A_{\mu\omega}^\dag A_{\nu\omega}
-
\Tr(A_{\mu\omega}^\dag A_{\nu\omega}\rhoSb^{(0)})
\big]
\\&+
\lambda^2\sum_{\mu,\nu}\sum_{\omega\neq\omega'}
G_{\mu\nu}(\omega,\omega')
A_{\mu\omega'}^\dag A_{\nu\omega}
\Big\}.
\end{split}
\end{equation}

\section{Hamiltonian of mean force}
\label{SecHMF}

Now we want to express $\rhoSb$ in the quasi-Gibbsian form as

\begin{equation}\label{EqHMF}
\rhoSb=\ZMF^{-1} e^{-\beta\HMF(\beta)},
\end{equation}
where 
\begin{equation}\label{EqHMFln}
\HMF(\beta)=-\frac1\beta\ln\left(\frac{\TrB e^{-\beta H}}{\ZB}\right)
\end{equation}
 is the Hamiltonian of mean force and $\ZMF=\Tr e^{-\beta\HMF(\beta)}=\ZSB/\ZB$. We want to find $\HMF$ in the second order in $\lambda$:
\begin{equation}\label{EqHMF2}
\HMF(\beta)\cong\HS+\lambda^2\HMF^{(2)}(\beta).
\end{equation}

We can do this if we compare expansion (\ref{EqMFGnonnorm}) with the corresponding expansion obtained from Eqs.~(\ref{EqHMF}) and (\ref{EqHMF2}). We again use the Kubo expansion (\ref{EqKubo}) with $H_0=\HS$ and $V=\HMF^{(2)}(\beta)$:
\begin{equation}
\exp\big\lbrace\!-\!\beta\big[\HS+\lambda^2\HMF^{(2)}(\beta)\big]\big\rbrace
\cong
e^{-\beta\HS}
\left(
1-\lambda^2\int_0^\beta 
e^{\beta_1\HS}\HMF^{(2)}(\beta)e^{-\beta_1\HS}\,d\beta_1
\right).
\end{equation}
Now we decompose
\begin{equation}
\HMF^{(2)}(\beta)=\sum_\omega H^{\rm MF,\,(2)}_\omega(\beta),
\qquad [H_S,H^{\rm MF}_\omega(\beta)]=
-\omega H^{\rm MF,\,(2)}_\omega(\beta),
\end{equation}
so that
\begin{equation}
\int_0^\beta 
e^{\beta_1\HS}\HMF^{(2)}(\beta)e^{-\beta_1\HS}\,d\beta_1
=\sum_{\omega}\frac{1-e^{-\beta\omega}}\omega
H^{\rm MF,\,(2)}_\omega(\beta)
\end{equation}
(the term with $\omega=0$ is special, but formally it can be obtained from the general one by evaluating the limit $\omega\to0$). Comparison with Eq.~(\ref{EqMFGnonnorm}) gives
\begin{equation}\label{EqHMFomega}
H^{\rm MF, \,(2)}_\omega(\beta)=
-\frac{\omega}{1-e^{-\beta\omega}}
\sum_{\mu,\nu}\sum_{\omega'-\omega''=\omega}
G_{\mu\nu}(\omega',\omega'')A_{\mu\omega''}^\dag A_{\nu\omega'}.
\end{equation}
So,
\begin{equation}
\HMF^{(2)}(\beta)=
-\sum_{\omega,\omega'}
\frac{\omega-\omega'}{1-e^{-\beta(\omega-\omega')}}
G_{\mu\nu}(\omega,\omega') A_{\mu\omega'}^\dag A_{\nu\omega}
\end{equation}
and
\begin{equation}\label{EqHMFresult}
\HMF(\beta)\cong
\HS-\lambda^2\sum_{\mu,\nu}\sum_{\omega,\omega'}
\frac{\omega-\omega'}{1-e^{-\beta(\omega-\omega')}}
G_{\mu\nu}(\omega,\omega') A_{\mu\omega'}^\dag A_{\nu\omega}.
\end{equation}

Now let us separate the terms with $\omega=\omega'$ and with $\omega\neq\omega'$:
\begin{equation}\label{EqHMFsep}
\HMF(\beta)\cong
\HS-
\lambda^2\sum_{\mu,\nu}\sum_{\omega}\beta^{-1}
G_{\mu\nu}(\omega,\omega) A_{\mu\omega}^\dag A_{\nu\omega}
-\lambda^2\sum_{\mu,\nu}\sum_{\omega\neq\omega'}
\frac{\omega-\omega'}{1-e^{-\beta(\omega-\omega')}}
G_{\mu\nu}(\omega,\omega') A_{\mu\omega'}^\dag A_{\nu\omega}.
\end{equation}
Since $[\HS,A_{\mu\omega}^\dag A_{\nu\omega}]=0$, the terms with $\omega=\omega'$ (the diagonal part) only modify the eigenvalues but not eigenvectors (unless this perturbation leads to degeneracy splitting). The terms with $\omega'\neq\omega$ (the off-diagonal part) mainly modify the eigenvectors. They also modify the eigenvalues, but, according to the general perturbation theory of quantum mechanics, give corrections of the order $O(\lambda^4)$.

Formulas (\ref{EqHMFresult}) and (\ref{EqHMFsep}) consitute a final result of this section. Note that $\HMF^{(2)}(\beta)$ is Hermitian, which can be easily checked.

\section{Hamiltonian of mean force in the exponential form}
\label{SecHMFexp}

Formulas (\ref{EqHMFresult}) and (\ref{EqHMFsep})  still can be transformed into the exponential form. See Ref.~\cite{ValkunasHMF} and Eqs.~(\ref{EqHMFhighValk}) and (\ref{EqHMFhighSingle}) for an analogous result for the high temperature approximation. From Eq.~(\ref{EqHMFsep}), let us define a system Hamiltonian with the diagonal correction
\begin{equation}\label{EqHSprime}
\HS'=\HS-\lambda^2\sum_{\mu,\nu}\sum_{\omega}\beta^{-1}
G_{\mu\nu}(\omega,\omega) A_{\mu\omega}^\dag A_{\nu\omega}.
\end{equation}
This Hamiltonian has its own spectrum and Bohr frequencies. Let us replace expansions (\ref{EqAomega}) and (\ref{EqHMFomega}) over the eigenoperators of $[\HS,\,\cdot\,]$ by the corresponding expansions over the eigenoperators of $[\HS',\,\cdot\,]$ (with the same notation). If the perturbation in Eq.~(\ref{EqHMFsep}) does not change the multiplicities of the eigenvalues, then decompositions (\ref{EqAomega}) and (\ref{EqHMFomega}) are not changed. The Bohr frequencies $\omega$ are changed on the order $O(\lambda^2)$. Since the corrections we discuss already have the second order, this infinitesimal change of Bohr frequencies has effect only in the higher orders.  Then, since
\begin{equation}
[\HS',A_{\mu\omega'}^\dag A_{\nu\omega}]=
-(\omega-\omega')A_{\mu\omega'}^\dag A_{\nu\omega},
\end{equation}
Eq.~(\ref{EqHMFsep}) can be rewritten as 
\begin{equation}
\HMF(\beta)\cong
\HS'+\lambda^2\sum_{\mu,\nu}\sum_{\omega\neq\omega'}
\frac{G_{\mu\nu}(\omega,\omega')}{1-e^{-\beta(\omega-\omega')}}
[\HS',A_{\mu\omega'}^\dag A_{\nu\omega}]
\cong\HMF^{\rm exp}(\beta),
\end{equation}
where
\begin{equation}\label{EqHMFexp}
\HMF^{\rm exp}(\beta)=e^{i\lambda^2 R}\HS' e^{-i\lambda^2R},
\end{equation}
\begin{equation}
R=i\sum_{\mu,\nu}\sum_{\omega,\omega'}
\frac{G_{\mu\nu}(\omega,\omega')}{1-e^{-\beta(\omega-\omega')}}
A_{\mu\omega'}^\dag A_{\nu\omega}.
\end{equation}
So, the correction to the isolated system Hamiltonian $\HS$ can be splitted into two parts: (i) corrections to the eigenvalues by the diagonal part of the perturbation (\ref{EqHSprime}), and (ii) a rotation of the eigenvectors with the generator $R$. Representation (\ref{EqHMFexp}) might be more convenient than (\ref{EqHMFresult})--(\ref{EqHMFsep}).

We just need to discuss the following issue. We have replaced decompositions (\ref{EqAomega}) and (\ref{EqHMFomega}) over the eigenoperators of $[\HS,\,\cdot\,]$ by the corresponding decompositions eigenoperators of $[\HS',\,\cdot\,]$, but then $\HS'$ itself also depends on this decomposition. In the simplest case, we can use the Hamiltonian $\HS$ in the definition of $\omega$ and $A_{\mu\omega}$ in Eq.~(\ref{EqHSprime}) and then switch to the decompositions with respect to $\HS'$. More complicated self-consistent schemes of definition of $\HS'$ and the decompositions also can be used.

\section{The case of harmonic oscillator bath}
\label{SecHarmonic}

We have obtained general formulas. In order to use them in practice, the coefficients $G_{\mu\nu}(\omega,\omega')$ should be evaluated. The simplest and widely studied case is the harmonic bath with the linear coupling (with respect to the bath ladder operators):

\begin{gather}
\HB=\sum_\xi\omega_\xi a^\dag_\xi a_\xi,\label{EqHB}\\
B_\mu=\sum_\xi \big(\overline{g_{\mu\xi}}a_\xi+g_{\mu\xi}a^\dag_\xi\big)
\label{EqB},
\end{gather}
where the index $\xi$ enumerates the oscillating modes and $g_{\mu\xi}$ are complex coupling constants.
Define the spectral densities:
\begin{equation}
\mathcal J_{\mu\nu}(\omega)
=\sum_{\xi}\overline{g_{\mu\xi}}g_{\nu\xi}\delta(\omega-\omega_\xi).
\end{equation}

Then, 
\begin{gather}
C_{\mu\nu}(\beta')=\int_0^\infty d\omega
\left\lbrace
\mathcal J_{\mu\nu}(\omega)[n_\beta(\omega)+1]e^{-\beta'\omega}
+
\mathcal J_{\nu\mu}(\omega)n_\beta(\omega)e^{\beta'\omega}
\right\rbrace,\label{EqCharm}
\\
\hat C_{\mu\nu}(\omega)=\int_0^\beta C_{\mu\nu}(\beta')e^{\beta'\omega}d\beta'
=-\left[S_{\mu\nu}(\omega)+e^{\beta\omega}S_{\nu\mu}(-\omega)\right],\label{EqCint}
\end{gather}
where
\begin{equation}\label{EqS}
S_{\mu\nu}(\omega)=
\int_0^\infty
d\omega_1
\left\{
\mathcal J^*_{\mu\nu}(\omega_1)
\frac{n_\beta(\omega_1)}{\omega_1+\omega}
-
\mathcal J_{\mu\nu}(\omega_1)
\frac{n_\beta(\omega_1)+1}{\omega_1-\omega}
\right\},
\end{equation}
$n_\beta(\omega)=[e^{\beta\omega}-1]^{-1}$ is the Bose-Einstein distribution. In the case of discrete oscillating modes in Eqs.~(\ref{EqHB}) and~(\ref{EqB}), all integrals over $\omega$ and $\omega_1$ are reduced to sums. In the thermodynamic limit of continuous modes (i.e., when the sums over $\xi$ in Eqs.~(\ref{EqHB}) and~(\ref{EqB}) are substituted by integrals), the integral in (\ref{EqS}) should be understood as a principal part integral.

\begin{remark}
It is interesting to note that $S_{\mu\nu}(\omega)$ are coefficients of the Lamb-shift Hamiltonian for the Davies (or secular Bloch-Redfield) master equation, which are usually defined as 
\begin{equation}\label{EqSLamb}
S_{\mu\nu}(\omega)=\frac1{2i}[\Gamma_{\mu\nu}(\omega)-
\Gamma^*_{\nu\mu}(\omega)],
\end{equation} 
where
\begin{equation}\label{EqGamma}
\Gamma_{\mu\nu}(\omega)=
\int_0^\infty C_{\mu\nu}(it)e^{it\omega}dt.
\end{equation}
This coincidence can be understood if we equivalently express the integral in (\ref{EqCint}) as an integral in the complex plane over the contour constituted by segments $\beta'\in[0,i\infty]$, $\beta'\in[i\infty,\beta+i\infty]$, and $\beta'\in[\beta+i\infty,\beta]$ provided that this contour integral exists. Then, the right-hand side of (\ref{EqCint}) with $S_{\mu\nu}$ defined by (\ref{EqSLamb})--(\ref{EqGamma}) also can be obtained. However, the direct calculation of the integral in (\ref{EqCint}) is more general since it is valid for a finite number of oscillators in the bath (even for a single oscillator), while the integral in Eq.~(\ref{EqGamma}) converges only in the thermodynamic limit of an infinite number of oscillators. See also Ref.~\cite{AlickiRenorm} for discussion of relations between the Lamb shift and the Hamiltonian of mean force.
\end{remark}

Then (see Ref.~\cite{LatuneReact}),
\begin{equation}
\begin{split}
G_{\mu\nu}(\omega,\omega')&=
\frac{1}{\omega'-\omega}
\left[
e^{\beta(\omega'-\omega)}
\hat C_{\mu\nu}(\omega)-\hat C_{\mu\nu}(\omega')
\right]
\\
&=
\frac{1}{\omega'-\omega}
\left\{
S_{\mu\nu}(\omega)
+e^{\beta\omega'}[S_{\nu\mu}(-\omega')-S_{\nu\mu}(-\omega)]
-e^{\beta(\omega'-\omega)}S_{\mu\nu}(\omega)
\right\}
\end{split}
\end{equation}
for $\omega\neq\omega'$ and
\begin{equation}
\begin{split}
G_{\mu\nu}(\omega,\omega)
&=\beta\hat C_{\mu\nu}(\omega)+
\hat C'_{\mu\nu}(\omega)
\\
&=S'_{\mu\nu}(\omega)-e^{\beta\omega}S'_{\mu\nu}(-\omega)-\beta S_{\mu\nu}(\omega).
\end{split}
\end{equation}
The derivatives are:
\begin{gather}
\hat C_{\mu\nu}'(\omega)=
-S'_{\mu\nu}(\omega)+S'_{\nu\mu}(-\omega)
-\beta e^{\beta\omega} S_{\nu\mu}(-\omega),\\
S'_{\mu\nu}(\omega)=-\int_0^\infty
d\omega_1
\left\{
\mathcal J^*_{\mu\nu}(\omega_1)
\frac{n_\beta(\omega_1)}{(\omega_1+\omega)^2}
+
\mathcal J_{\mu\nu}(\omega_1)
\frac{n_\beta(\omega_1)+1}{(\omega_1-\omega)^2}
\right\}.
\end{gather}
Explicit formulas for the cases over- and underdamped oscillators were derived in Ref.~\cite{LatuneReact}. 

\section{High-temperature approximation}
\label{SecHigh}

In Ref.~\cite{ValkunasHMF}, an expression for the mean-force Hamiltonian in the high-temperature was derived for a specific model of excitation energy transfer. Here we generalize these results to the general case of a system coupled to a thermal bath of harmonic oscillators.  

Now we expand (\ref{EqMFGgen}) in the powers of $\beta$ (rather than $\lambda$) up to the second order. Note that $C_{\mu\nu}(\omega)$ in Eq.~(\ref{EqCharm}) is of the order $\beta^{-1}$:
\begin{equation}
n_\beta(\omega)e^{\beta'\omega}
\cong
\frac{1}{\beta\omega}
+\frac{\beta'}{\beta}-\frac12,
\qquad
[n_\beta(\omega)+1]e^{-\beta'\omega}
\cong
\frac{1}{\beta\omega}
-\frac{\beta'}{\beta}+\frac12,
\end{equation}
and
\begin{equation}\label{EqCapprox}
C_{\mu\nu}(\beta')\cong\frac{2\Re\Lambda_{\mu\nu}}\beta
+
i\left(
1-\frac{2\beta'}\beta
\right)
\Im M_{\mu\nu},
\end{equation}
where 
\begin{equation}\label{EqLambda}
\Lambda_{\mu\nu}
=
\int d\omega\,
\frac{\mathcal J_{\mu\nu}(\omega)}{\omega},
\qquad
M_{\mu\nu}=\int d\omega\,\mathcal J_{\mu\nu}(\omega).
\end{equation}
$\Lambda_{\mu\mu}$ correspond to reorganization energies in the theory of excitation energy transfer. Thus, to obtain the term of the order $\beta^2$, we should consider the fourth-order term with respect to $\HI$ in Eq.~(\ref{EqMFGgen}):

\begin{equation}\label{EqMFGgen4}
\begin{split}
\TrB e^{-\beta H}=
\ZB\,e^{-\beta\HS}
\bigg[
1&+
\int_0^{\beta}d\beta_1
\int_0^{\beta_1}d\beta_2\,
\langle
\HI(\beta_1)\HI(\beta_2)
\rangle_{{\rm B},\beta}
\\
&+
\int_0^{\beta}d\beta_1
\int_0^{\beta_1}d\beta_2
\int_0^{\beta_2}d\beta_3
\int_0^{\beta_3}d\beta_4\,
\langle
\HI(\beta_1)\HI(\beta_2)\HI(\beta_3)\HI(\beta_4)
\rangle_{{\rm B},\beta}
\bigg].
\end{split}
\end{equation}
Here we put $\lambda=1$ since now it is not small  and we do not need this formal parameter. 

To calculate the quadruple correlator, we use Wick's theorem:

\begin{equation}\label{EqMFGHighT}
\begin{split}
\TrB e^{-\beta H}\cong\ZB\,e^{-\beta\HS}
\bigg\lbrace
1+
\sum_{\mu,\nu}
\int_0^\beta d\beta_1
\int_0^{\beta_1} d\beta_2\,
&C_{\mu\nu}(\beta_2)
A_\mu^\dag(\beta_1)A_\nu(\beta_2)
\\
+
\sum_{\mu_1,\nu_1,\mu_1,\nu_2}
\int_0^{\beta}d\beta_1
\int_0^{\beta_1}d\beta_2
\int_0^{\beta_2}d\beta_3
\int_0^{\beta_3}d\beta_4\,
&C_{\mu_1\nu_1}(\beta_1-\beta_2)C_{\mu_2\nu_2}(\beta_3-\beta_4)
\\
\times
\big[&A_{\mu_1}^\dag (\beta_1)
A_{\nu_1}(\beta_2)
A_{\mu_2}^\dag(\beta_3)
A_{\nu_2}(\beta_4)
\\
+
&A_{\mu_1}^\dag (\beta_1)
A_{\mu_2}^\dag(\beta_2)
A_{\nu_1}(\beta_3)
A_{\nu_2}(\beta_4)
\\
+
&A_{\mu_1}^\dag (\beta_1)
A_{\mu_2}^\dag(\beta_2)
A_{\nu_2}(\beta_3)
A_{\nu_1}(\beta_4)\big].
\end{split}
\end{equation}
Since the quadruple integral gives the order $\beta^4$, we should consider only the $\beta^{-2}$-part of the integrand: $C_{\mu\nu}(\beta')\cong2\Re\Lambda_{\mu\nu}/\beta$, $A_\mu(\beta')\cong A_\mu$, and

\begin{equation}
\begin{split}
\int_0^{\beta}&d\beta_1
\int_0^{\beta_1}d\beta_2
\int_0^{\beta_2}d\beta_3
\int_0^{\beta_3}d\beta_4\,
\langle
\HI(\beta_1)\HI(\beta_2)\HI(\beta_3)\HI(\beta_4)
\rangle_{{\rm B},\beta}
\\
&\cong
\frac{\beta^2}6
\sum_{\mu_1,\nu_1,\mu_1,\nu_2}
\Re\Lambda_{\mu_1\nu_1}\Re\Lambda_{\mu_2\nu_2}
(A_{\mu_1}^\dag A_{\nu_1}A_{\mu_2}^\dag A_{\nu_2}
+
A_{\mu_1}^\dag A_{\mu_2}^\dag A_{\nu_1} A_{\nu_2}
+
A_{\mu_1}^\dag A_{\mu_2}^\dag A_{\nu_2} A_{\nu_1}).
\end{split}
\end{equation}

Now consider the double integral expression in Eq.~(\ref{EqMFGHighT}). We use Eq.~(\ref{EqCapprox}) and 
\begin{equation}
A_\mu(\beta')\cong A_\mu+\beta'[\HS,A_\mu].
\end{equation}
Since the double integration gives the order $\beta^2$, we should consider the terms proportional to $\beta^{-1}$ and $\beta^0$ in the integrand:
\begin{equation}
\begin{split}
&\int_0^\beta d\beta_1
\int_0^{\beta_1} d\beta_2\,
C_{\mu\nu}(\beta_2)
A_\mu^\dag(\beta_1)A_\nu(\beta_2)
\\
&\cong
\int_0^\beta d\beta_1
\int_0^{\beta_1} d\beta_2
\left[
\frac{2\Re\Lambda_{\mu\nu}}\beta
\left(
A_\mu^\dag A_\nu
+\beta_1[\HS,A_\mu^\dag]A_\nu
+\beta_2A_\mu^\dag[\HS,A_\nu]
\right)
+i\Im M_{\mu\nu} 
\left(1-\frac{2\beta_2}{\beta}\right)
A_\mu^\dag A_\nu
\right]
\\
&=
\Re\Lambda_{\mu\nu}
\left(
\beta A_\mu^\dag A_\nu
+\frac{2\beta^2}3[\HS,A_\mu^\dag]A_\nu
+\frac{\beta^2}3A_\mu^\dag[\HS,A_\nu]
\right)
+\frac{i\beta^2}6\Im M_{\mu\nu}A_\mu^\dag A_\nu.
\end{split}
\end{equation}

In order to express the result in the quasi-Gibbsian form (\ref{EqHMF}) and to find the Hamiltonian of mean force, we first express the corrections to the Gibbs state in the exponential form (with the condition that the expansion up to the second order in $\beta$ should coincide with the expressions obtained above):
\begin{equation}
\begin{split}
\TrB e^{-\beta H}\cong\ZB\,
e^{-\beta\HS}
\exp\bigg\lbrace
&\sum_{\mu,\nu}\Re\Lambda_{\mu\nu}
\left(
\beta A_\mu^\dag A_\nu
+\frac{2\beta^2}3[\HS,A_\mu^\dag]A_\nu
+\frac{\beta^2}3A_\mu^\dag[\HS,A_\nu]
\right)
\\
+
&\frac{i\beta^2}6\Im M_{\mu\nu}A_\mu^\dag A_\nu
-\frac{\beta^2}3
\sum_{\mu_1,\nu_1,\mu_2,\nu_2}
\Re\Lambda_{\mu_1\nu_1}\Re\Lambda_{\mu_2\nu_2}
A_{\mu_1}^\dag A_{\nu_1}A_{\mu_2}^\dag A_{\nu_2}
\\
+&\frac{\beta^2}6
\sum_{\mu_1,\nu_1,\mu_2,\nu_2}
\Re\Lambda_{\mu_1\nu_1}\Re\Lambda_{\mu_2\nu_2}
A_{\mu_1}^\dag A_{\mu_2}^\dag\{A_{\nu_1},A_{\nu_2}\}
\bigg\rbrace
\end{split}
\end{equation}

Now we can use the Baker-Campbell-Hausdorff formula:
\begin{equation}
e^{-\beta\HS}e^{\beta T}=
\exp\left(-\beta\HS+\beta T-\frac{\beta^2}2[\HS,T]+O(\beta^3)
\right),
\end{equation}
so that
\begin{equation}
\begin{split}
\TrB e^{-\beta H}\cong\ZB
\exp\bigg\lbrace
\!&\!-\!\beta\HS
+\!\sum_{\mu,\nu}\Re\Lambda_{\mu\nu}
\left(
\beta A_\mu^\dag A_\nu
\!+\!\frac{2\beta^2}3[\HS,A_\mu^\dag]A_\nu
\!+\!\frac{\beta^2}3A_\mu^\dag[\HS,A_\nu]
\!-\!\frac{\beta^2}2[\HS,A_\mu^\dag A_\nu]
\right)
\\
&+\sum_{\mu,\nu}\frac{i\beta^2}6\Im M_{\mu\nu}A_\mu^\dag A_\nu
-\frac{\beta^2}3
\sum_{\mu_1,\nu_1,\mu_2,\nu_2}
\Re\Lambda_{\mu_1\nu_1}\Re\Lambda_{\mu_2\nu_2}
A_{\mu_1}^\dag A_{\nu_1}A_{\mu_2}^\dag A_{\nu_2}
\\
&+\frac{\beta^2}6
\sum_{\mu_1,\nu_1,\mu_2,\nu_2}
\Re\Lambda_{\mu_1\nu_1}\Re\Lambda_{\mu_2\nu_2}
A_{\mu_1}^\dag A_{\mu_2}^\dag\{A_{\nu_1},A_{\nu_2}\}
\bigg\rbrace
\end{split}
\end{equation}
and the Hamiltonian of mean force is
\begin{equation}\label{EqHMFhigh}
\begin{split}
\HMF&\cong\HS-\sum_{\mu,\nu}\Re\Lambda_{\mu\nu}A_\mu^\dag A_\nu
+\frac{\beta}{3}
\sum_{\mu,\nu}\Re\Lambda_{\mu\nu}
\left(
A_\mu^\dag\HS A_\nu
-
\frac12
\{\HS,A_\mu^\dag A_\nu\}
\right)
\\
&-\sum_{\mu,\nu}\frac{i\beta}6\Im M_{\mu\nu}A_\mu^\dag A_\nu
+\frac{\beta}3
\sum_{\mu_1,\nu_1,\mu_2,\nu_2}
\Re\Lambda_{\mu_1\nu_1}\Re\Lambda_{\mu_2\nu_2}
A_{\mu_1}^\dag A_{\nu_1}A_{\mu_2}^\dag A_{\nu_2}
\\
&-\frac{\beta}6
\sum_{\mu_1,\nu_1,\mu_2,\nu_2}
\Re\Lambda_{\mu_1\nu_1}\Re\Lambda_{\mu_2\nu_2}
A_{\mu_1}^\dag A_{\mu_2}^\dag\{A_{\nu_1},A_{\nu_2}\}.
\end{split}
\end{equation}
The last term in the first line of Eq.~(\ref{EqHMFhigh}) has the form of the Franke--Gorini--Kossakowski--Lindblad--Sudarshan (FGKLS) generator.\footnote{The abbreviations GKLS or GKSL are more common \cite{L,GKS,BriefGKLS}; however, in the same 1976, this general form of a generator of a quantum dynamical semigroup was also proposed by V.\,A.\,Franke \cite{Franke}, see also Ref.~\cite{Andrianov}.} Since $\beta$ is a small parameter, we can substitute its sum with $H_S$ by the action of the semigroup $\Phi_{\beta/3}$ with this generator. This is a generalization of the double exponentiation in Ref.~\cite{ValkunasHMF}:
\begin{equation}\label{EqHMFhighPhi}
\begin{split}
\HMF&\cong\Phi_{\beta/3}(\HS)-\sum_{\mu,\nu}\Re\Lambda_{\mu\nu}A_\mu^\dag A_\nu
+
\\
&-\sum_{\mu,\nu}\frac{i\beta}6\Im M_{\mu\nu}A_\mu^\dag A_\nu
+\frac{\beta}3
\sum_{\mu_1,\nu_1,\mu_2,\nu_2}
\Re\Lambda_{\mu_1\nu_1}\Re\Lambda_{\mu_2\nu_2}
A_{\mu_1}^\dag A_{\nu_1}A_{\mu_2}^\dag A_{\nu_2}
\\
&-\frac{\beta}6
\sum_{\mu_1,\nu_1,\mu_2,\nu_2}
\Re\Lambda_{\mu_1\nu_1}\Re\Lambda_{\mu_2\nu_2}
A_{\mu_1}^\dag A_{\mu_2}^\dag\{A_{\nu_1},A_{\nu_2}\}.
\end{split}
\end{equation}

Let us now show how the formula of Ref.~\cite{ValkunasHMF} can be obtained as a particular case from Eq.~(\ref{EqHMFhighPhi}). Let $g_{\mu\xi}g_{\nu\xi}=\delta_{\mu\nu} g_{\mu\xi}^2$ and $A_{\mu}=\ket\mu\bra\mu$ with $\{\ket\mu\}$ being an orthonormal basis. Then the last two lines in Eqs.~(\ref{EqHMFhigh}) and (\ref{EqHMFhighPhi}) vanish. Also, 
\begin{equation}\label{EqHMFhighSimp}
\Phi_{\beta/3}(\HS)=\HS^{\rm diag}+
e^{-\beta\hat\Lambda/6}\HS^{\text{off-diag}}e^{-\beta\hat\Lambda/6},
\end{equation}
where $\hat\Lambda=\sum_\mu\Lambda_\mu\ket\mu\bra\mu$ is the reorganization Hamiltonian and $\HS^{\text{diag}}$ and $\HS^{\text{off-diag}}$ are the diagonal and off-diagonal part of the system Hamiltonian $\HS$ in the basis $\{\ket\mu\}$. So,
\begin{equation}\label{EqHMFhighValk}
\HMF\cong\HS^{\rm diag}-\hat\Lambda+
e^{-\beta\hat\Lambda/6}\HS^{\text{off-diag}}e^{-\beta\hat\Lambda/6},
\end{equation}
which is exactly the result of Ref.~\cite{ValkunasHMF}. The last term expresses the suppression of the off-diagonal part of the system Hamiltonian due to non-vanishing interaction with the bath. In the theory of excitation energy transfer, the states $\ket\mu$ correspond to local site excitations. So, the interaction with the bath reduces the intersite couplings, which lead to a well-known effect of more localized eigenvectors (excitons).

Let us consider another particular case. Let the interaction Hamiltonian (\ref{EqHI}) has only a single term, so that the sums over $\mu$ and $\nu$ disappear. This again leads to vanishing of the last two lines in Eqs.~(\ref{EqHMFhigh}) and (\ref{EqHMFhighPhi}) and formula (\ref{EqHMFhighSimp}). Let, further, the spectrum of $A$ be non-degenerate: $A=\sum_n a_n\ket n$. Then,
\begin{equation}\label{EqHMFhighSingle}
\HMF\cong\HS^{\rm diag}-\Lambda A^2+
\sum_{n\neq m}J_{nm}\,e^{-\beta\Lambda(a_n-a_m)^2/6}\ket n\bra m,
\end{equation}
where $\HS^{\rm diag}$ is the diagonal part of the system Hamiltonian in the basis $\{\ket n\}$ and $J_{nm}=\braket{n|\HS|m}$. Again, the interaction with the bath suppresses the off-diagonal elements of the system Hamiltonian in the eigenbasis of the interaction Hamiltonian.

It can be seen that formulas (\ref{EqHMFhighValk}) and (\ref{EqHMFhighSingle}) correctly reproduce the Hamiltonian of mean force in the ultrastrong-coupling limit $\Lambda\to\infty$ (or $\lambda\to\infty$ if we restore the dimensionless coupling constant $\lambda$) for all temperatures obtained in Ref.~\cite{CresserAnders}, see also Refs.~\cite{LatuneUltrastrong,TrushUltra,Keeling}. In particular, it is diagonal in the so called ``pointer basis'' (the eigenbasis of $A$ in the case of Eq.~(\ref{EqHMFhighSingle}) and the eigenbasis $\{\ket\mu\}$ of all $A_\mu$ in the case of Eq.~(\ref{EqHMFhighValk})).

General formulas (\ref{EqHMFhigh}) and (\ref{EqHMFhighPhi}) and formula (\ref{EqHMFhighSingle}) for a new particular case are the main results of this section.

\section{Simulation}
\label{SecSim}

For the simulation, let us consider a qubit coupled to a ``bath'' of a single harmonic oscillator:

\begin{equation}\label{EqSingleOsc}
\HS=\frac\varepsilon2\sigma_z,
\qquad
\HB=\Omega a^\dag a,
\qquad
\HI=\frac{\sigma_z-\sigma_x}{\sqrt2}\otimes g(a+a^\dag).
\end{equation}
Here $\sigma_z=\ket1\bra1-\ket0\bra0$ and $\sigma_x=\ket1\bra0+\ket0\bra1$ are the Pauli matrices.

\begin{figure}[h]
\begin{centering}
\includegraphics[width=\textwidth]{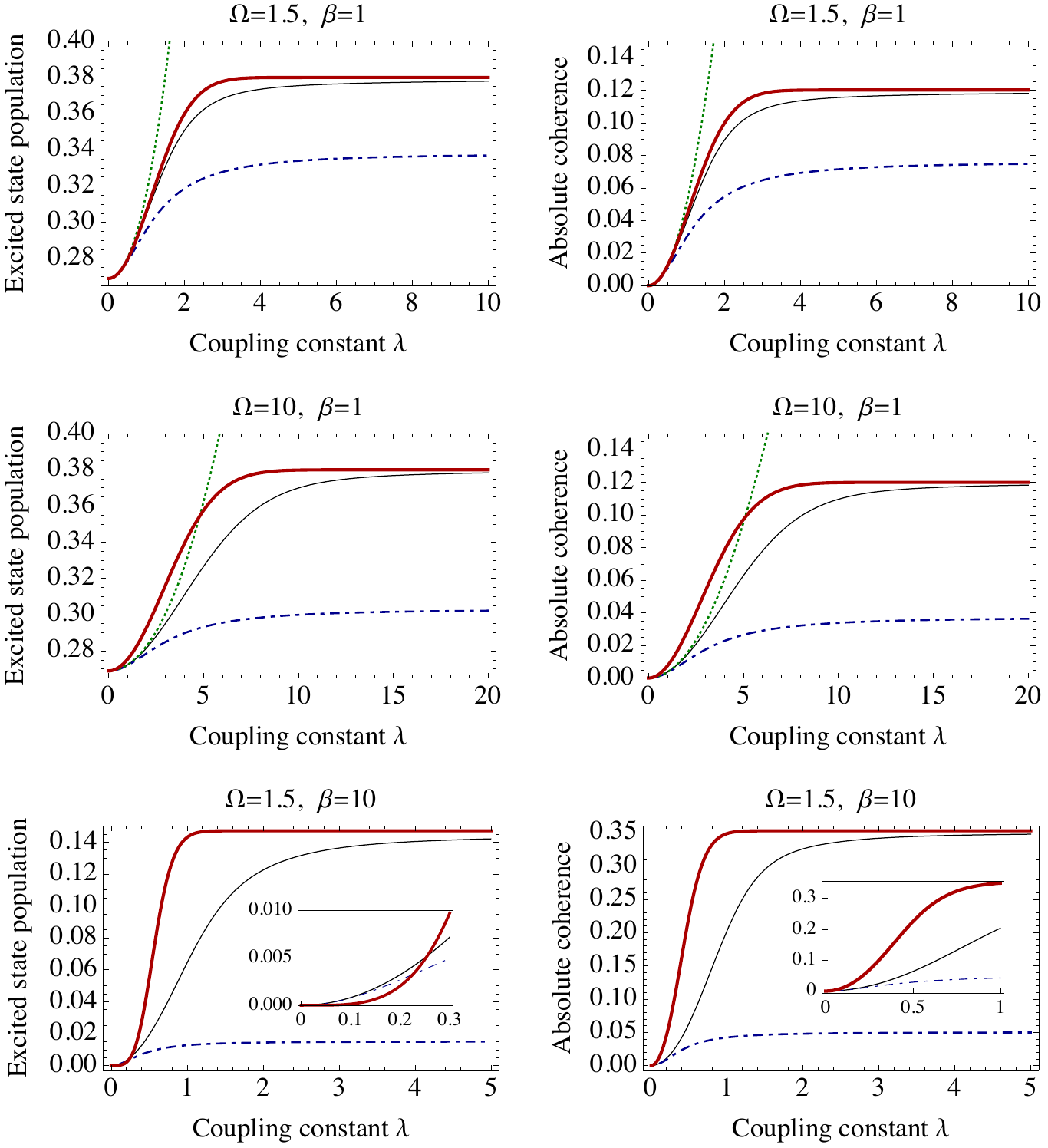}
%\vskip -2mm
\caption{\small
Excited state population $\braket{1|\rhoSb|1}$ and the absolute coherence $|\braket{0|\rhoSb|1}|$ for various $\Omega$ and $\beta$ depending on the coupling constant $\lambda$ for model (\ref{EqSingleOsc}) of a qubit coupled to a ``bath'' of a single harmonic oscillator with $\varepsilon=1$. Black line: Numerically exact method; Blue dash-dotted line: Direct approximation of  the mean force Gibbs state in the weak-coupling regime (\ref{EqMFG}); Green dotted line: Formula (\ref{EqHMF}) with approximation (\ref{EqHMFresult}) of the Hamiltonian of mean force in the weak-coupling regime; Red thick solid line: Formula (\ref{EqHMF}) with approximation (\ref{EqHMFhighSingle}) of the Hamiltonian of mean force in the high-temperature regime. In the last row,  approximation (\ref{EqHMF}) gives large errors for almost all $\beta$ and not shown. In the first two lines, the green dotted line goes outside the plot area, which also means that approximation (\ref{EqHMF}) gives large errors for the corresponding $\beta$. Inset in the last line: a part of the plots for small $\lambda$ in more detail.
}
\label{FigSingleOscL}
\end{centering}
\end{figure}

\begin{figure}[h]
\begin{centering}
\includegraphics[width=\textwidth]{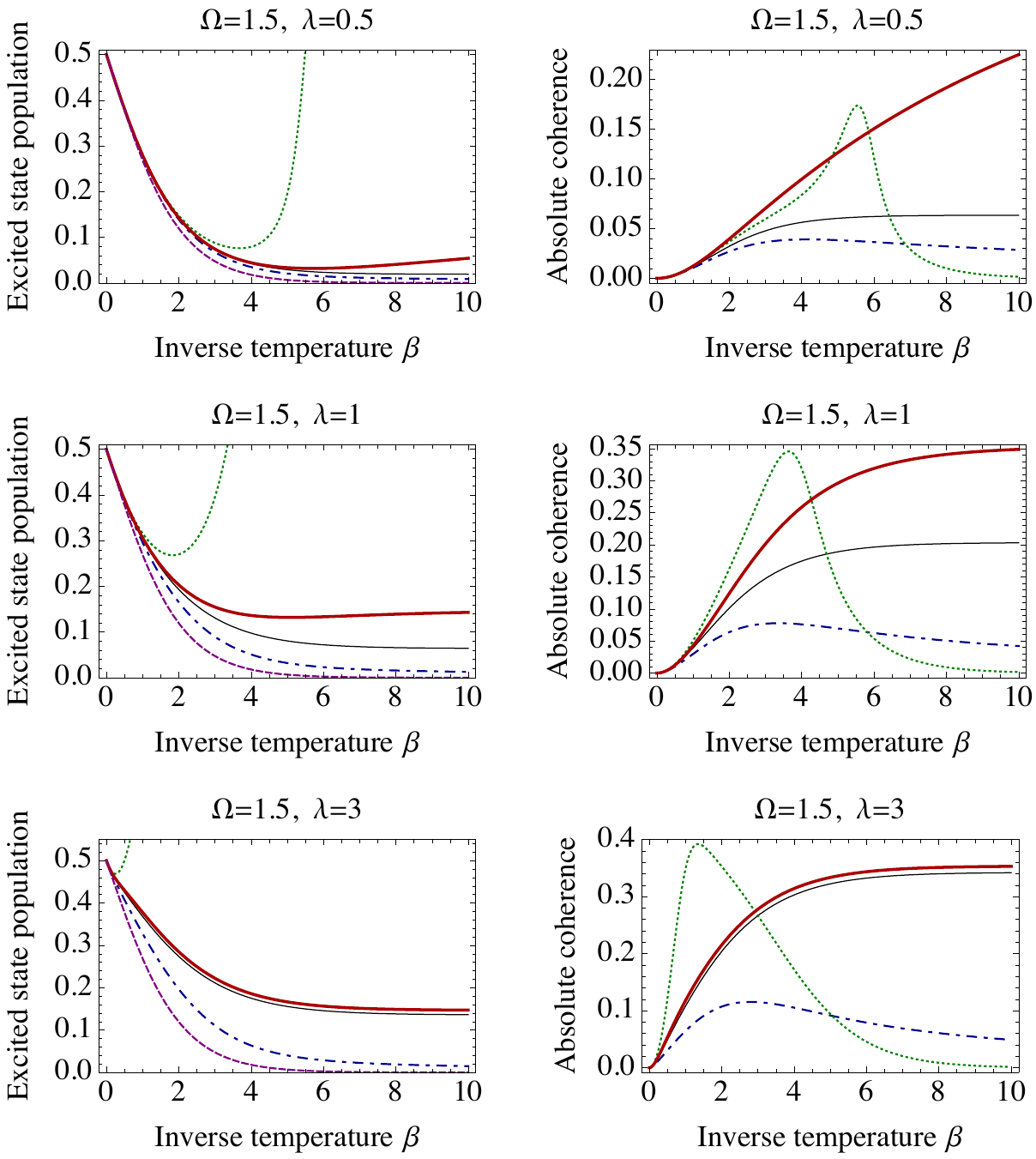}
%\vskip -2mm
\caption{\small
Excited state population $\braket{1|\rhoSb|1}$ and the absolute coherence $|\braket{0|\rhoSb|1}|$ for various coupling constants $\lambda$ depending on the inverse temperature $\beta$ for model (\ref{EqSingleOsc}) of a qubit coupled to a ``bath'' of a single harmonic oscillator with $\varepsilon=1$ and $\Omega=1.5$. Different colours of lines correspond to various approximations, see Fig.~\ref{FigSingleOscL}. The additional purple dashed lines correspond to the Gibbs state $\rhoSb^{(0)}\propto e^{-\beta\HS}$.
}
\label{FigSingleOscB1}
\end{centering}
\end{figure}

\begin{figure}[h]
\begin{centering}
\includegraphics[width=\textwidth]{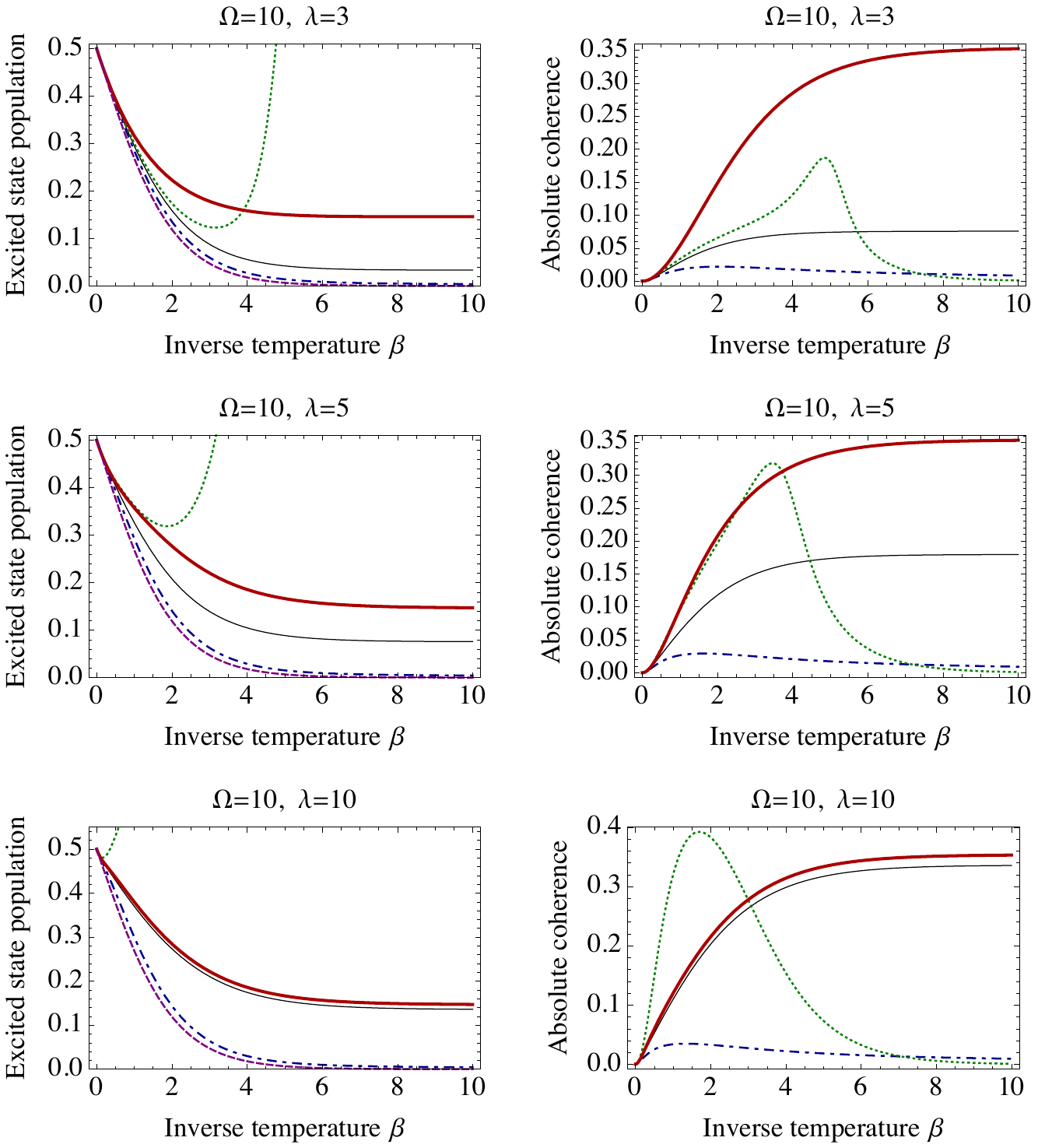}
%\vskip -2mm
\caption{\small
Excited state population $\braket{1|\rhoSb|1}$ and the absolute coherence $|\braket{0|\rhoSb|1}|$ for various coupling constants $\lambda$ depending on the inverse temperature $\beta$ for model (\ref{EqSingleOsc}) of a qubit coupled to a ``bath'' of a single harmonic oscillator with $\varepsilon=1$ and $\Omega=10$. The lines correspond to various approximations, see Fig.~\ref{FigSingleOscL}. The additional purple dashed lines correspond to the Gibbs state $\rhoSb^{(0)}\propto e^{-\beta\HS}$.
}
\label{FigSingleOscB2}
\end{centering}
\end{figure}

We put $\varepsilon=1$. In other words, all units are measured in terms of $\varepsilon$, $\varepsilon^{-1}$, or the degrees of $\varepsilon$, depending on the dimensionality. For this model, the mean force Gibbs state can be calculated in a numerically exact way: We can truncate the energy levels of the oscillator much larger than $\beta\Omega$ and then numerically take the partial trace of the corresponding finite-dimensional system-bath Gibbs state. We compare the approximate calculations of the mean force Gibbs state according to: (i) the direct approximation of  the mean force Gibbs state in the weak-coupling regime (\ref{EqMFG}), (ii) formula (\ref{EqHMF}) with approximation (\ref{EqHMFresult}) of the Hamiltonian of mean force in the weak-coupling regime, and (iii) formula (\ref{EqHMF}) with approximation (\ref{EqHMFhighSingle}) of the Hamiltonian of mean force in the high-temperature regime with (iv) the numerically exact result. We do not show the performance of  approximation (\ref{EqHMFexp}) for the Hamiltonian of mean force in the weak-coupling regime in the exponential form since it always gives slightly less precise results than the mean-force Hamiltonian in the original form (\ref{EqHMFresult}). So, the exponential form is interesting mainly from the conceptual point of view. Plots of the excited state population $\braket{1|\rhoSb|1}$ and the absolute coherence $|\braket{0|\rhoSb|1}|$ for various $\Omega$ and $\beta$ are given on Figs.~\ref{FigSingleOscL}--\ref{FigSingleOscB2}.

If we compare the two approximations for the weak-coupling perturbation theory, we see that both approximations work for small $\lambda$ and $\beta$ (large temperatures). However, the direct approximation of the mean-force Gibbs state (\ref{EqMFG}) is more reliable that the approximation by means of the Hamiltonian of mean force (\ref{EqHMFresult}). If the coupling constant $\lambda$ is small, the latter does not work for small temperatures, while the former works well for all considered temperatures. Also, in some cases, the direct approximation provides reasonable results even for large $\lambda$. 

The high-temperature approximation works well in the case of high-temperature and also in the case of large coupling constants even for low temperatures. The reason for the last fact is that this approximation correctly reproduces the ultrastrong-coupling limit, as noticed above.

\section{Refined Bloch-Redfield quantum master equation}
\label{SecRef}

The Bloch-Redfield quantum master equation is a widely used equation describing the dynamics of the reduced density operator of the system $\rho(t)$ in the weak-coupling regime \cite{BP,RH}. It can be written in the following GKLS-like form \cite{FarinaGio,CattaneoLocGlob}:

\begin{equation}\label{EqRedfieldGKLS}
\dot\rho(t)=-i[H_{\rm LS},\rho(t)]
+\sum_{\mu,\nu}
\gamma_{\mu\nu}(\omega,\omega')
\Big(A_{\nu\omega}\rho(t)A_{\mu\omega'}^\dag-
\frac12\big\{A_{\mu\omega'}^\dag A_{\nu\omega},\rho(t)\big\}
\Big),\qquad
\end{equation}
where
\begin{equation}\label{EqHLSRedf}
H_{\rm LS}=\sum_{\omega,\omega'}\sum_{\mu,\nu}
S_{\mu\nu}(\omega,\omega')
A_{\mu\omega'}^\dag A_{\nu\omega},
\end{equation}
(the subindex LS stands for the Lamb shift),
\begin{equation}
\begin{split}
\gamma_{\mu\nu}(\omega,\omega')&=
\Gamma_{\mu\nu}(\omega)+\Gamma^*_{\nu\mu}(\omega'),\\
S_{\mu\nu}(\omega,\omega')&=
\frac1{2i}\left[\Gamma_{\mu\nu}(\omega)-\Gamma^*_{\nu\mu}(\omega')
\right]
\end{split}
\end{equation}
(cf. Eq.~\ref{EqSLamb}), where $\Gamma_{\mu\nu}(\omega)$ is defined in Eq.~(\ref{EqGamma}). The terms with $\omega\neq\omega'$ oscillate with the frequencies $|\omega-\omega'|$ in the interaction picture. If these frequencies are much greater than the rate of evolution of $\rho(t)$, then we can neglect them. This is called the secular approximation. The secular Bloch-Redfield quantum master equation has the FGKLS form. Its right-hand side is also called the Davies generator \cite{MerkliRev}. In the case of the thermal bath, it turns out that
\begin{equation}
\gamma_{\mu\nu}(-\omega,-\omega)
=e^{-\beta\omega}\gamma_{\nu\mu}(\omega,\omega)
\end{equation}
and, as a consequence, the Gibbs state $\rhoSb^{(0)}$ is a stationary solution of  the master equation. In the generic case, this stationary soulution is unique and any solution of the master equation converges to it on large times.

In Refs.~\cite{KM,KMcorr,MerkliRev}, it is proposed to formally replace the system Hamiltonian $\HS$ by the Hamiltonian of mean force $\HMF$ in the construction of the master equation. So, decomposition (\ref{EqAomega}) over the eigenoperators of $[\HS,\,\cdot\,]$ is replaced by the corresponding decomposition with respect to the eigenoperators of $[\HMF,\,\cdot\,]$. Then, the solution of the master equation converges to the exact steady state $\rhoSb$. We can hope that the new (``refined'') quantum master equation at least does not lose the precision on intermediate times. Unfortunately, rigorous results of  Refs.~\cite{KM,KMcorr,MerkliRev} does not guarantee the last point. It would be interesting to compare the numerical solutions of the corresponding equations with a numerically exact method. As the latter, we choose the hierarchical equations of motion (HEOM) in the high-temperature approximation \cite{IFl}.

\begin{figure}[h]
\begin{centering}
\includegraphics[width=\textwidth]{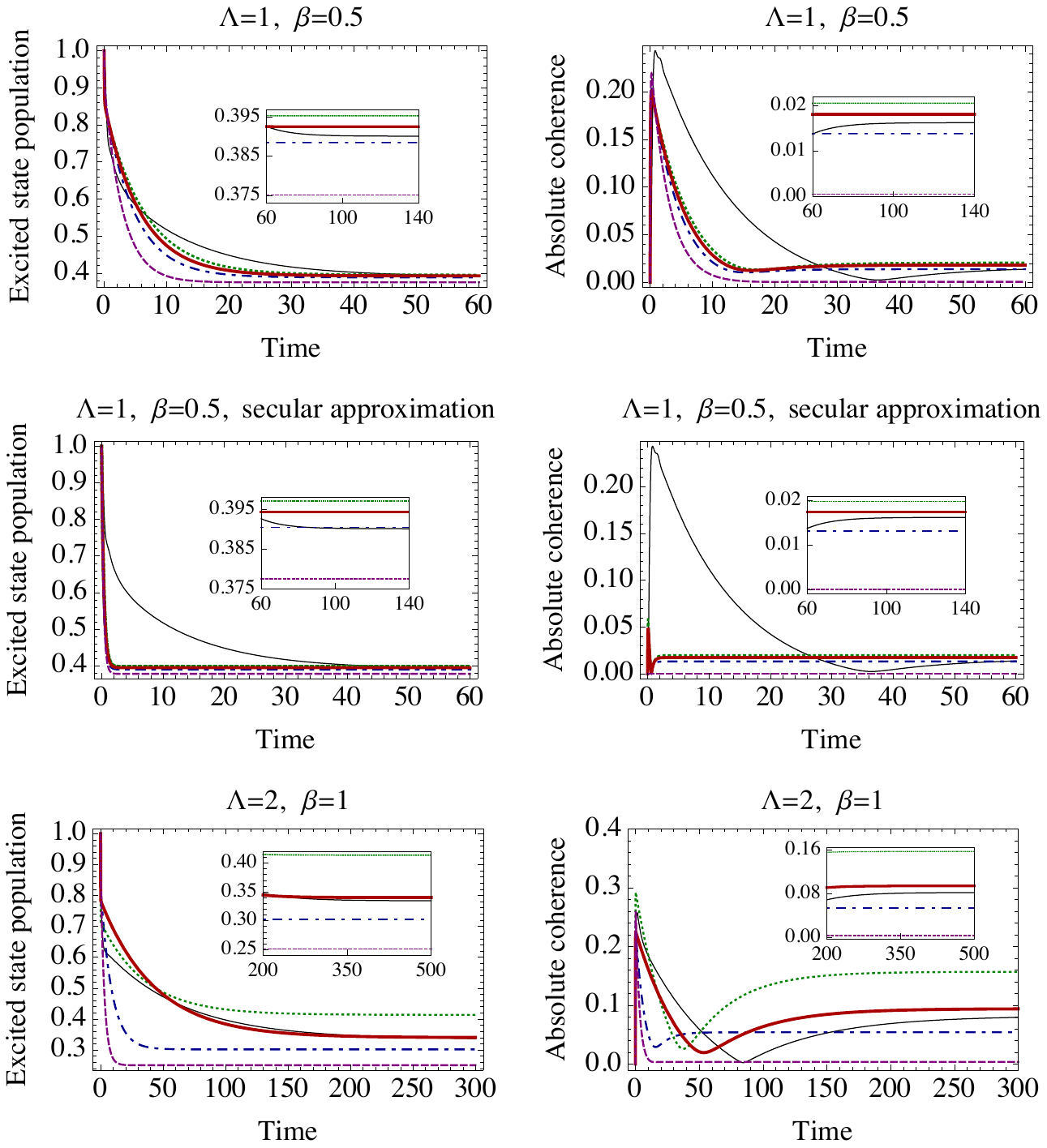}
%\vskip -2mm
\caption{\small
Performance of the Bloch-Redfield equation with the system Hamiltonian replaced by the Hamiltonian of mean force for different coupling constants and temperatures. The second row shows the performance of the secular Bloch-Redfield equation, while the other rows shows the performance of the full Bloch-Redfield equation. Different curves correspond to different approximations for the Hamiltonian of mean force. Black line: Numerically exact method of HEOM; Blue dash-dotted line: Approximation (\ref{EqMFGnonnorm}) and (\ref{EqHMFln}); Green dotted line: Weak-coupling approximation (\ref{EqHMFresult}); Red thick solid line: High-temperature approximation (\ref{EqHMFhighSingle}). Insets: long-time behaviour (steady-state solutions).
}
\label{FigRefined}
\end{centering}
\end{figure}

Comparison of numerical solutions of the original Bloch-Redfield equation (both full and secular) and the refined Bloch-Redfield equation with various approximations of $\HMF$ is given on Fig.~\ref{FigRefined}. We tested approximations (\ref{EqHMFresult}), (\ref{EqHMFexp}), and (\ref{EqHMFhighSimp}) for $\HMF$ and also the approximation based on  formula (\ref{EqMFGnonnorm}) for the (non-normalized) mean-force Gibbs state. If we have a direct approximation of the mean-force Gibbs state like Eq.~(\ref{EqMFGnonnorm}), then we can approximate the Hamiltonian of mean force as well using the formula (\ref{EqHMFln}).

For the simulation, we take the following model:
\begin{equation}
\HS=\frac\varepsilon2,\quad 
\HB=\int \omega(\xi)a(\xi)^\dag a(\xi)\,d\xi,
\quad
\HI=\frac{\sigma_z-\sigma_x}{\sqrt2}
\otimes\int[\overline{g(\xi)}a(\xi)+g(\xi)a(\xi)^\dag]\,d\xi,
\end{equation}
and the Drude--Lorentz spectral density for the bath (the overdamped oscillator model):
\begin{equation}
\mathcal J(\omega)=\frac{\Lambda}\pi
\frac{2\omega\gamma}{\omega^2+\gamma^2}.
\end{equation}
It turns out that
\begin{equation}
\int_0^\infty d\omega\,\frac{\mathcal J(\omega)}{\omega}=\Lambda,
\end{equation}
which is consistent with definition of the reorganization energy $\Lambda$ (\ref{EqLambda}). As the initial system-bath state, we take
\begin{equation}\label{EqIni}
\rho^{\rm tot}(0)=\ket1\bra 1\otimes\rhoBb.
\end{equation}
In our simulation, again, $\varepsilon=1$ (i.e., all other quantities can be considered to be expressed in terms of $\varepsilon$, $\varepsilon^{-1}$, etc.) and also $\gamma=0.5$. The high-temperature approximation for the HEOM works if $\beta\gamma<1$ \cite{IFl}. The formal parameter $\lambda$ is equal to one. It can be treated to be absorbed in the interaction Hamiltonian, so that $\Lambda$ is proportional to $\lambda^2$. The role of the system-bath coupling parameter is played then by the reorganization energy $\Lambda$, which is common, e.g., in the theory of excitation energy transfer.

From the plots, we see that the replacement of the system Hamiltonian by the Hamiltonian of mean force indeed improves the precision of the solutions of the Bloch-Redfield quantum master equation on both large and intermediate times. The short-term behaviour is not shown in detail because this is not a focus here. The peculiarities of the short-term behaviour are caused by the initial product state (\ref{EqIni}). During the short initial time, the product state evolves into a state adjusted to the system-bath dynamics \cite{Slip,TrushBog,Taepra}. To improve the short-term performance, one can consider the time-dependent version of the Bloch-Redfield equation (replace the infinite upper limit of integration in Eq.~(\ref{EqGamma}) by the current time instant $t$) or the cumulant approach \cite{Alicki,RivasRef,RivasRefTD,WinczewskiBypassing}.

We see that, for the chosen parameters, the secular Bloch-Redfield equation does not work well. The secular approximation works well for smaller values of $\Lambda$, but then the correction to the Gibbs state $\rhoSb-\rhoSb^{(0)}$, which is the focus of the present paper, is also smaller. Note that, in the weak-coupling regime, the secular approximation may fail due to nearly degenerate energy levels or Bohr frequencies. In this case, certain non-secular generalizations of the Davies master equation of the FGKLS form (mostly based on different kinds of partial secular approximations) are proposed \cite{CattaneoLocGlob,FarinaGio,PtaEsp,NathanRudner,Uni}. However, here we cannot use such modifications here because there are no quasi-degenerate energy levels or Bohr frequencies here. Strictly speaking, we consider the parameters beyond the weak-coupling regime, but the Bloch-Redfield equation (especially, the refined one, with $\HS$ replaced by $\HMF$) still gives reasonable results since it contains the minimal number of assumptions in comparison with the FGKLS equations for the weak-coupling regime.

The stationary solution of the full (non-secular) Bloch-Redfield equation is not exactly the Gibbs state. In particular, it does not commute with $\HS$ (or with $\HMF$ in the case of the refined master equation). Moreover, the coherences (off-diagonal elements in the eigenbasis of $\HS$) of the steady state coincide with those of $\rhoSb$ in the principal order \cite{FlemingCummings,Tupkary}. This is considered to be an advantage of the full Bloch-Redfield equation over the secular one. However, this is not an advantage if we replace $\HS$ by $\HMF$ in order to have $\rhoSb$ as the exact stationary solution of a quantum master equation. The stationary state of the refined full Bloch-Redfield equation slightly differs from $\rhoSb$. However, this is not a major drawback since certain error already originates from the approximate expressions for $\HMF$. From the plots, we see that, for the chosen parameters, the refined full Bloch-Redfield equation is approximately as good as the refined secular Bloch-Redfield equation on large times and much better on intermediate times.

Finally, we see that the refinement of the master equation based on approximation (\ref{EqHMFresult}) for the Hamiltonian of mean force in the weak-coupling approximation becomes less precise for larger coupling constants and smaller temperatures, while approximations (\ref{EqMFGnonnorm}) and (\ref{EqHMFhighSimp}) still give quite precise predictions of the steady state. Approximation (\ref{EqMFGnonnorm}) works better if both the coupling constant (reorganization energy) and the temperature are small and approximation (\ref{EqHMFhighSimp}) works better if either the coupling constant or the temperature is large.

\section{Conclusions}

The main results of this work is the formulas (\ref{EqHMFresult}), (\ref{EqHMFsep}) and (\ref{EqHMFexp}) for the Hamiltonian of mean force in the weak-coupling regime and formulas (\ref{EqHMFhighPhi}) and (\ref{EqHMFhighSingle}) for the Hamiltonian of mean force in the  high-temperature approximation. From numerical estimations, we see that precision of all these approximations (as well as the direct approximation (\ref{EqMFG}) for the mean force Gibbs state) depends on various parameters: coupling constant, temperature, parameters of the bath spectral density. In particular, we see that the high-temperature approximation works well even for low temperatures whenever the coupling constant is large enough. So, it is worthwhile to find analytical estimates for the errors of these approximations depending on all these parameters.

Also we have shown that the formal replacement of the system Hamiltonian by the Hamiltonian of mean force in the construction of the quantum master equation can improve the performance of the master equation on intermediate and large times.

\textbf{Acknowledgements.} We are grateful to Alexander Teretenkov, Camille Lombard Latune, Janet Anders, James Cresser, Marcin {\L}obejko, Marco Merkli, and Marek Winczewski for ongoing fruitful discussions on corrections to the Gibbs state. This work is supported by the Russian Science Foundation under grant 17-71-20154.


\begin{thebibliography}{99}

\bibitem{TMCA}
A.\,S.\,Trushechkin, M.\,Merkli, J.\,D.\,Cresser, and J.\,Anders, Open quantum system dynamics and the mean force Gibbs state, \textit{AVP Quantum Science} (to appear), {\href{https://arxiv.org/abs/2110.01671}{arXiv:2110.01671}}.

\bibitem{Grabert1984}
H.\,Grabert, U.\,Weiss, and P.\,Talkner,
Quantum theory of the damped harmonic oscillator,
\href{https://doi.org/10.1007/BF01307505}{{\em Z. Phys. B} {\bf 55}(4), 87--94 (1984)}.

\bibitem{Subasi}
Y.\,Suba\c{s}i, C.\,H.\,Fleming, J.\,M.\,Taylor, and B.\,L.\,Hu,
Equilibrium states of open quantum systems in the strong coupling regime,
\href{https://doi.org/10.1103/PhysRevE.86.061132}
{{\em Phys. Rev. E} {\bf 86}(6), 061132 (2012)}.


\bibitem{Geva}
E.\,Geva, E.\,Rosenman, and D.\,Tannor,
On the second-order corrections to the quantum canonical equilibrium density matrix,
\href{https://doi.org/10.1063/1.481928}{{\em J. Chem. Phys.} {\bf 113}(4), 1380--1390 (2000)}.



\bibitem{Mori}
T.\,Mori and S.\,Miyashita, Dynamics of the density matrix in contact with a thermal bath and the quantum master equation,
\href{https://doi.org/10.1143/JPSJ.77.124005}
{{\em J. Phys. Soc. Jpn.} {\bf 77}(12), 124005 (2008)}.

\bibitem{Thingna}
J.\,Thingna, J.-S.\,Wang, and P.\,H\"anggi, 
Generalized Gibbs state with modified Redfield solution: Exact agreement up to second order,
\href{https://doi.org/10.1063/1.4718706}{{\em J. Chem. Phys.} {\bf 136}(19), 194110 (2012)}.



\bibitem{Purkaqubit}
A.\,Purkayastha, G.\,Guarnieri, M.\,T.\,Mitchison, R.\,Filip, and J.\,Goold,
Tunable phonon-induced steady-state coherence in a double-quantum-dot charge qubit,
\href{https://doi.org/10.1063/1.4718706}{{\em npj Quantum Inf.} {\bf 6}, 27 (2020)}.
 

\bibitem{CresserAnders}
J.\,D.\,Cresser and J.\,Anders, Weak and ultrastrong coupling limits of the quantum mean force Gibbs state, 
\href{https://doi.org/10.1103/PhysRevLett.127.250601}{{\em Phys. Rev. Lett.} {\bf 127}(25), 250601 (2021)}.


\bibitem{LatuneReact}
C.\,L.\,Latune,
Steady state in strong bath coupling: reaction coordinate versus perturbative expansion,
{\href{https://arxiv.org/abs/2110.03169}{arXiv:2110.03169}}.

\bibitem{LatuneUltrastrong}
C.\,L.\,Latune,
Steady state in ultrastrong coupling regime: perturbative expansion and first orders,
{\href{https://arxiv.org/abs/2110.02186}{arXiv:2110.02186}}.

\bibitem{Cao2012jcp}
C.-K.\,Lee, J.\,Moix, and J.\,Cao,
Accuracy of second order perturbation theory in the polaron and variational polaron frames,
\href{https://doi.org/10.1063/1.4722336}
{{\em J. Chem. Phys.} {\bf 136}(20), 204120 (2012)}.

\bibitem{Cao2012pre}
C.-K.\,Lee, J.\,Cao, and J.\,Gong,
Noncanonical statistics of a spin-boson model: Theory and exact Monte Carlo simulations,
\href{https://doi.org/10.1103/PhysRevE.86.021109}
{{\em Phys. Rev. E} {\bf 86}(2), 021109 (2012)}.

\bibitem{Cao2016}
D.\,Xu and J.\,Cao, 
Non-canonical distribution and non-equilibrium transport beyond weak system-bath coupling regime: A polaron
transformation approach,
\href{https://doi.org/10.1007/s11467-016-0540-2}
{{\em Front. Phys.} {\bf 11}(4), 110308 (2016)}.


\bibitem{ValkunasHMF}
A. Gelzinis and L. Valkunas, Analytical derivation of equilibrium state for open quantum system,
\href{https://doi.org/10.1063/1.5141519}{{\em J. Chem. Phys.} {\bf 152}(5), 051103 (2019)}.

\bibitem{GelinThoss}
M.\,F.\,Gelin and M.\,Thoss,
Thermodynamics of a subensemble of a canonical ensemble,
\href{https://doi.org/10.1103/PhysRevE.79.051121}
{{\em Phys. Rev. E} {\bf 79}(5), 051121 (2009)}.

\bibitem{Jarzynski}
C.\,Jarzynski,
Stochastic and Macroscopic Thermodynamics of Strongly Coupled Systems,
\href{https://doi.org/10.1103/PhysRevX.7.011008}
{{\em Phys. Rev. X} {\bf 7}(1), 011008 (2017)}.

\bibitem{MillerAnders}
H.\,J.\,D.\,Miller and J.\,Anders,
Entropy production and time asymmetry in the presence of strong interactions,
\href{https://doi.org/10.1103/PhysRevE.95.062123}
{{\em Phys. Rev. E} {\bf 95}(6), 062123 (2017)}.

\bibitem{MillerHMF}
H. J. D. Miller, 
Hamiltonian of mean force for strongly-coupled systems,
in 
\href{https://doi.org/10.1007/978-3-319-99046-0_22}
{\textit{Thermodynamics in the Quantum Regime: Fundamental Aspects and New Directions}, edited by F.\,Binder, L.\,A.\,Correa, C.\,Gogolin, J.\,Anders, and G.\,Adesso (Springer International Publishing, Cham, 2018), pp. 531--549}.

\bibitem{StrasbergEsposito}
P.\,Strasberg and M.\,Esposito, 
Measurability of nonequilibrium thermodynamics in terms of the Hamiltonian of mean force,
\href{https://doi.org/10.1103/PhysRevE.101.050101}
{{\em Phys. Rev. E} {\bf 101}(5), 050101(R) (2020)}.


\bibitem{Rivas}
\'A.\,Rivas,
Strong coupling thermodynamics of open quantum systems,
{\href{https://doi.org/10.1103/PhysRevLett.124.160601}{\textit{Phys. Rev. Lett}. {\bf 124}(16), 160601 (2020)}}.


\bibitem{TalknerHanggi}
P.\,Talkner and P.\,H\"anggi,
\textit{Colloquium}: Statistical mechanics and thermodynamics at strong coupling: Quantum and classical,
\href{https://doi.org/10.1103/RevModPhys.92.041002}
{{\em Rev. Mod. Phys.} {\bf 92}(4), 041002 (2020)}.




\bibitem{TereEffGibbs}
A.\,E.\,Teretenkov,
Effective Gibbs state for averaged observables,
{\href{https://arxiv.org/abs/2110.14407}{arXiv:2110.14407}}.


\bibitem{KM}
M.\,K\"onenberg and M.\,Merkli,
Completely positive dynamical semigroups and quantum resonance theory,
\href{https://doi.org/10.1007/s11005-017-0937-z}
{{\em Lett. Math. Phys.} {\bf 107}, 1215--1233 (2017)}.

\bibitem{KMcorr}
M.\,K\"onenberg and M.\,Merkli,
Correction to: Completely positive dynamical semigroups and quantum resonance theory,
\href{https://doi.org/10.1007/s11005-019-01177-9}
{{\em Lett. Math. Phys.} {\bf 109}, 1701--1702 (2019)}.

\bibitem{MerkliRev}
M.\,Merkli, 
Quantum Markovian master equations: Resonance theory shows validity for all time scales,
{\href{http://dx.doi.org/10.1016/j.aop.2019.167996}{\textit{Ann. Phys.} \textbf{412}, 167996 (2020)}}.   

\bibitem{BP}
H.-P.\,Breuer and F.\,Petruccione, 
{\it The Theory of Open Quantum Systems} 
(Oxford University Press, Oxford, 2002).

\bibitem{RH}
A.\,Rivas and S.\,F.\,Huelga,
{\it Open Quantum Systems: An introduction}
(Springer, Berlin, 2012).


\bibitem{AccLuVol}
L.\,Accardi, Y.\,G.\,Lu, and I.\,Volovich, \textit{Quantum Theory and Its Stochastic Limit}  (Springer, Berlin, 2002).

\bibitem{AccPechVol}
L.\,Accardi, A.\,N.\,Pechen, and I.\,V.\,Volovich, 
Quantum stochastic equation for the low density limit, 
\href{https://doi.org/10.1088/0305-4470/35/23/306}
{\textit{J. Phys. A} \textbf{35}(23), 4889--4902 (2002)}.

\bibitem{PechVol}
A.\,N.\,Pechen and I.\,V.\,Volovich, Quantum multipole noise and generalized quantum stochastic equations, 
\href{https://doi.org/10.1142/S0219025702000857}
{Infin. Dimens. Anal. Quantum Probab. Relat. Top. \textbf{5}(4), 441--464 (2002).}

\bibitem{AVK}
I.\,Ya.\,Aref'eva, I.\,V.\,Volovich, and S.\,V.\,Kozyrev, Stochastic limit method and interference in quantum many-particle systems, 
\href{https://doi.org/10.1007/s11232-015-0296-9}
{\textit{Theoret. and Math. Phys.} \textbf{183}(3), 782--799 (2015)}.


\bibitem{VolFunc}
I.\,V.\,Volovich, Functional mechanics and time irreversibility problem, in \href{https://doi.org/10.1142/9789814304061_0033}{\textit{Quantum bio-informatics III, QP-PQ: Quantum Probab. White Noise Anal.} \textbf{26} (World Sci. Publ., Hackensack, 2010), 393--404}.

\bibitem{VolTrush2009}
A.\,S.\,Trushechkin and I.\,V.\,Volovich, Functional classical mechanics and rational numbers, 
\href{https://doi.org/10.1134/S2070046609040086}
{\textit{$p$-Adic Numbers, Ultrametric Anal. Appl.} {\bf 1}(4), 361--367 (2009)}.

\bibitem{VolBog}
I.\,V.\,Volovich, Bogoliubov equations and functional mechanics, 
\href{https://doi.org/10.1007/s11232-010-0090-7}
{{\em Theoret. and Math. Phys.} {\bf 164}(3), 1128--1135 (2010)}.

\bibitem{VolPisk}
E.\,V.\,Piskovskiy and I.\,V.\,Volovich, On the correspondence between newtonian and functional mechanics, in
\href{https://doi.org/10.1142/9789814343763_0028}
{\textit{Quantum bio-informatics IV: From quantum information to bio-informatics, QP-PQ Quantum Probability and White Noise Analysis} \textbf{28}, (World Sci. Publ., Hackensack, 2011), 363--372}.

\bibitem{VolFuncStoch}
I.\,V.\,Volovich, Functional stochastic classical mechanics, 
\href{https://doi.org/10.1134/S2070046615010057}
{\textit{$p$-Adic Numbers Ultrametric Anal. Appl.} \textbf{7}(1), 56-70 (2015)}.

\bibitem{VolHoloTherm}
I.\,Ya.\,Aref'eva and I.\,V.\,Volovich, 
Holographic thermalization, 
\href{https://doi.org/10.1007/s11232-013-0016-2}
{\textit{Theoret. and Math. Phys.} \textbf{174}(2), 186--196 (2013)}.


\bibitem{TrushVolLocGlob}
A.\,S.\,Trushechkin and I.\,V.\,Volovich, Perturbative treatment of inter-site couplings in the local description of open quantum networks, 
\href{https://doi.org/10.1209/0295-5075/113/30005}
{\textit{EPL}, \textbf{113}(3), 30005 (2016)}.

\bibitem{InoVol}
I.\,V.\,Volovich and O.\,V.\,Inozemcev, On the Thermalization Hypothesis of Quantum States, 
\href{https://doi.org/10.1134/S0081543821020255}
{\textit{Proc. Steklov Inst. Math.} \textbf{313}, 268--278 (2021)}.


\bibitem{AlickiRenorm}
M.\,Winczewski and R.\,Alicki,
Renormalization in the theory of open quantum systems
via the self-consistency condition,
{\href{https://arxiv.org/abs/2112.11962}{arXiv:2112.11962}}.


\bibitem{CattaneoLocGlob}
M.\,Cattaneo, G.\,L.\,Giorgi, S.\,Maniscalco, and R.\,Zambrini, 
Local versus global master equation with common and separate baths: superiority of the global approach in partial secular approximation,
{\href{https://doi.org/10.1088/1367-2630/ab54ac}{New J. Phys. \textbf{21}, 113045 (2019)}}.

\bibitem{FarinaGio}
D.\,Farina and V.\,Giovannetti, 
Open-quantum-system dynamics: Recovering positivity of the Redfield equation via the partial secular approximation,
{\href{https://doi.org/10.1103/PhysRevA.100.012107}{Phys. Rev. A \textbf{100}(1), 012107 (2019)}}.

\bibitem{GKS}
V.\,Gorini, A.\,Kossakowski, and E.\,C.\,G.\,Sudarshan, 
Completely positive dynamical semigroups of $N$-level systems,
{\href{https://doi.org/10.1063/1.522979}{J. Math. Phys. \textbf{17}, 821--825 (1976)}}.

\bibitem{L}
G.\,Lindblad, 
On the generators of quantum dynamical semigroups,
{\href{https://doi.org/10.1007/BF01608499}{Commun. Math. Phys. \textbf{48}, 119--130 (1976)}}.

\bibitem{BriefGKLS}
D.\,Chru\'{s}ci\'{n}ski and S.\,Pascazio,
A brief history of the GKLS equation, {\href{https://doi.org/10.1142/S1230161217400017}{Open Sys. Inf. Dyn. \textbf{24}, 1740001 (2017)}}.

\bibitem{Franke}
V.\,A.\,Franke, On the general form of the dynamical transformation of density matrices,
{\href{https://doi.org/10.1007/BF01051230}{Theoret. and Math. Phys. \textbf{72}, 406--413 (1976)}}.

\bibitem{Andrianov}
A.\,A.\,Andrianov, M.\,V.\,Ioffe, and O.\,O.\,Novikov, 
Supersymmetrization of the Franke-Gorini-Kossakowski-Lindblad-Sudarshan equation,
{\href{https://doi.org/10.1142/S1230161217400017}{J. Phys. A \textbf{52}, 425301 (2019)}}.


\bibitem{TrushUltra}
A.\,Trushechkin,
Quantum master equations and steady states for the ultrastrong-coupling limit and the strong-decoherence limit,
{\href{https://arxiv.org/abs/2109.01888}{arXiv:2109.01888}}.


\bibitem{Keeling}
Y.-F.\,Chiu, A.\,Strathearn, J.\,Keeling,
Numerical evaluation and robustness of the quantum mean force Gibbs state,
{\href{https://arxiv.org/abs/2112.08254}{arXiv:2112.08254}}.

\bibitem{IFl}
A.\,Ishizaki and G.\,R.\,Fleming, Unified treatment of quantum coherent and incoherent hopping dynamics in electronic energy transfer: Reduced hierarchy equation approach, {\href{http://dx.doi.org/10.1063/1.3155372}{\textit{J. Chem. Phys.} \textbf{130}(23), 234111 (2009)}}.

\bibitem{Slip}
A. Su\'{a}rez, R. Silbey, and I. Oppenheim, ``Memory effects in the relaxation of quantum open systems,'' \href{https://doi.org/10.1063/1.463831}{\textit{J. Chem. Phys.} \textbf{97}(7), 5101--5107 (1992)}.

\bibitem{TrushBog}
A.\,S.\,Trushechkin, Derivation of the Redfield quantum master equation and corrections to it by the Bogoliubov method, {\href{https://doi.org/10.1134/S008154382102022X}{\textit{Proc. Steklov Inst. Math.} {\bf 313}, 246--257 (2021)}}; extended version: arXiv:2108.03128.

\bibitem{Taepra}
A. E. Teretenkov, ``Non-perturbative effects in corrections to quantum master equation arising in Bogolubov-van Hove limit,'' \href{https://doi.org/10.1088/1751-8121/ac0201}{\textit{J. Phys. A} \textbf{54}(26), 265302 (2021)}.

\bibitem{Alicki}
R.\,Alicki,
Master equations for a damped nonlinear oscillator and the validity of the Markovian approximation,
{\href{https://doi.org/10.1103/PhysRevA.40.4077}{\textit{Phys. Rev. A} \textbf{40}(7), 4077 (1989)}}.

\bibitem{RivasRef}
\'A.\,Rivas, 
Refined weak-coupling limit: Coherence, entanglement, and non-Markovianity,
{\href{https://doi.org/10.1103/PhysRevA.95.042104}{\textit{Phys. Rev. A} \textbf{95}(4), 042104 (2017)}}.

\bibitem{RivasRefTD}
\'A.\,Rivas, 
Quantum thermodynamics in the refined weak coupling limit,
{\href{https://doi.org/10.3390/e21080725}{\textit{Entropy} \textbf{21}(8), 725 (2019)}}.

\bibitem{WinczewskiBypassing}
M.\,Winczewski, A.\,Mandarino, M.\,Horodecki, and R.\,Alicki,
Bypassing the intermediate times dilemma for open quantum system,
{\href{https://arxiv.org/abs/2106.05776}{arXiv:2106.05776}}.

\bibitem{PtaEsp}
K.\,Ptaszy\'nski and M.\,Esposito, 
Thermodynamics of quantum information flows,
{\href{https://doi.org/10.1103/PhysRevLett.122.150603}{Phys. Rev. Lett. \textbf{122}, 150603 (2019)}}.


\bibitem{Bondar}
G.\,McCauley, B.\,Cruikshank, D.\,I.\,Bondar, and K.\,Jacobs, 
Accurate Lindblad-form master equation for weakly damped quantum systems across all regimes,
{\href{https://doi.org/10.1038/s41534-020-00299-6}{npj Quantum Inf. \textbf{6}, 74 (2020)}}.


\bibitem{NathanRudner}
F.\,Nathan and M.\,S.\,Rudner,
Universal Lindblad equation for open quantum systems,
{\href{https://doi.org/10.1103/PhysRevB.102.115109}{Phys. Rev. B \textbf{102}, 115109 (2020)}}.

\bibitem{Uni}
A.\,Trushechkin, Unified Gorini-Kossakowki-Lindblad-Sudarshan quantum master equation beyond the secular approximation, {\href{https://doi.org/10.1103/PhysRevA.103.062226}{\textit{Phys. Rev. A} {\bf 103}(6), 062226 (2021)}}.

\bibitem{FlemingCummings}
C.\,H.\,Fleming and N.\,I.\,Cummings,
Accuracy of perturbative master equations,
{\href{https://doi.org/10.1103/PhysRevE.83.031117}{\textit{Phys. Rev. E} \textbf{83}(2), 031117 (2011)}}.

\bibitem{Tupkary}
D.\,Tupkary, A.\,Dhar, M.\,Kulkarni, and A.\,Purkayastha,
Fundamental limitations in Lindblad descriptions of systems weakly coupled to baths,
{\href{https://arxiv.org/abs/2105.12091}{arXiv:2105.12091}}.

\end{thebibliography}
\end{document}